%% file: cl.tex
\documentclass[12pt]{article}
\usepackage{fullpage}
\usepackage{amsmath}
\usepackage{amssymb}
\usepackage{epsfig}
\usepackage{feynmf}
\usepackage{subfigure}
\begin{document}
\begin{flushright}
ZU-TH 27/00
\end{flushright}

\begin{center}

\vspace{0.5cm}

{\Large \bf $\boldsymbol{ c\to u\gamma}$ in the minimal supersymmetric 
standard model }\\
\vspace{1cm}
S. Prelovsek

\vspace{.3cm}

{\it Department of Theoretical Physics, University of Trieste, Strada Costiera 
11, Italy

J. Stefan Institute, Jamova 39, 1000 Ljubljana, Slovenia}

 \vspace{.5cm}

and

\vspace{0.5cm}

D. Wyler

\vspace{0.3cm}

{\it Institut f\" ur Theoretische Physik, Universit\" at Z\" urich, Switzerland
}

\end{center}

\vspace{1cm}

\centerline{\bf ABSTRACT}

\vspace{0.3cm}

The minimal supersymmetric standard model (MSSM) with universal soft breaking 
terms has a negligible effect on the rare $c\to u\gamma$ decay rate. We 
therefore study a general model with non-universal soft breaking terms
constrained by vacuum stability and the experimental upper bound on the mass 
difference of the neutral charm mesons.  It turns out that gluino exchange 
can enhance the  
standard model rate by up to two orders of magnitudes; in contrast, the 
contributions from 
charginos and neutralinos remain at least an order of magnitude below 
the QCD corrected standard model rate.

\vspace{2cm}

{\bf 1. Introduction}

\vspace{0.2cm}

Flavour changing neutral currents (FCNC) occur in the standard model only 
at the loop level where they are suppressed by the GIM mechanism. 
The corresponding processes are rare and  suitable as probes for
new physics, in particular for the popular low energy supersymmetry (SUSY)
which has various new sources of flavour 
violation. Studying the FCNC processes may reveal the flavour 
structure of the soft supersymmetry breaking terms and shed light on the 
mechanism of SUSY breaking. There is a renewed interest in this 
issue, because models with effective supergravities \cite{supergravities} 
generally do not lead to universal soft breaking terms and tend to have
large FCNC rates. 

In this note we explore the sensitivity of the $c\to u\gamma$ decay rate 
in the framework of the  minimal supersymmetric standard model (MSSM) 
with unbroken R-parity and arbitrary non-universal soft breaking 
terms. This process was considered before in ref.
\cite{masiero.cugamma}; however,
only gluino exchanges were included and the constraints from vacuum stability
were not applied. In addition, we take into account the  QCD corrections within the standard model, which are seizable,  and discuss the QCD corrections in the MSSM. Motivated by recent proposals for 
probing $c\to u\gamma$ in specific hadronic observables 
\cite{FPSW,FPS,BFO,bigi}, 
which are largely free of the 'uninteresting' long distance contributions, 
we analyze this decay in more detail.

In the standard 
model, the $c\to u\gamma$ amplitude is proportional to $\sum_{q=d,s,b} 
V_{cq}^*V_{uq}m_q^2$ at the one-loop level and the corresponding branching 
ratio $\sim 10^{-17}$ is negligibly 
small as the GIM mechanism is extremely effective in charm decays 
(the intermediate $d$ and $s$ quarks are very light compared to the $W$).
The most important QCD corrections arise from the diagrams shown 
in Fig. 1. They drastically lift the GIM suppression; instead of power-like
it becomes logarithmic \cite{greub}. The effective local interaction 
can be written as \cite{greub} 
\begin{align}
\label{c7.sm}
{\cal L}_{eff}&=-\tfrac{4G_F}{\sqrt{2}}V_{cs}^*V_{us}~[c_7~{\cal O}_7+
c_7^\prime 
~{\cal O}_7^\prime]~\\
{\cal O}_7&=\tfrac{e}{16\pi^2}m_c~\bar u 
\sigma_{\mu\nu}\tfrac{1}{2}(1+\gamma_5)c~F^{\mu\nu}~,\qquad {\cal 
O}_7^\prime=\tfrac{e}{16\pi^2}m_c~\bar u 
\sigma_{\mu\nu}\tfrac{1}{2}(1-\gamma_5)c~F^{\mu\nu}\nonumber\\
(c_{7})_{\alpha_s}^{SM}&\simeq 
\frac{\alpha_s(m_c)}{4\pi}c_2(m_c)\{f[m_s^2/m_c^2]-f[m_d^2/m_c^2]\}\simeq 
-(0.007+0.020~ i) [1\pm 0.2]~\nonumber\\
(c_{7}^{\prime})^{SM}&\simeq 0\nonumber 
\end{align}
for $m_c=1.5$ GeV and $\alpha_s(m_c)=0.31$, with functions 
 $f$ and $c_2$  given in \cite{greub}. The corresponding branching ratio in the
 standard model 
\begin{equation}
\label{br.sm}
Br^{SM}_{\alpha_s}(c\to u\gamma)=6~ \biggl\vert {eV_{cs}^*V_{us}\over 
2\pi}\biggr\vert^2\bigl\{|c_7|^2+|c_7^\prime|^2\bigr\}~{G_F^2m_c^5\over 192\pi^
3 
\Gamma(D^0)}\simeq 3\times 10^{-8}
\end{equation}
is still small and potentially sensitive to various extensions beyond the 
standard model.

\input{fig/graf1.tex}

The main generic problem for probing the $c\to u\gamma$ transition in hadron 
decays is that the most straightforward observables such as the decay rate
$D\to \rho \gamma$ are largely dominated by
long distance contributions and  are independent of new physics 
\cite{radiative}. However, there are observables which are
dominated by the interesting short distance physics. For instance, the 
disturbing long 
distance contributions largely cancel in the difference of the decay rates 
$Br(D^0\to 
\rho^0\gamma)-Br(D^0\to \omega\gamma)$  as shown in \cite{FPSW}. Other 
possibilities to look for $c\to u\gamma$ are 
the $B_c\to B_u^*\gamma$ decay \cite{FPS}  and the violation of the equalities 
$Br(D_s^+\to K^{*+}\gamma)/Br(D_s^+\to \rho^+\gamma)=\tan\theta_C$ \cite{BFO}, 
$Br(D^0\to \rho^0\gamma)/Br(D^0\to \bar K^{*0}\gamma)=\tan\theta_C/2$ 
\cite{masiero.cugamma,bigi}. 
Similarly, the isospin violating quantity $\Gamma(D^+\to 
\rho^+\gamma)/2\Gamma(D^0\to \rho^0\gamma)-1$  \cite{ahl} might help to 
disentangle the
different contributions. These observables are suitable for testing  new 
scenarios that could significantly enhance the $c\to u\gamma$ rate, and 
are of particular interest in the context of SUSY flavour problem. \\

\input{fig/graf2.tex}

{\bf 2. The MSSM amplitudes }

\vspace{0.2cm}

The one-loop diagrams for $c\to u\gamma$ transition in MSSM are shown in 
Fig. 2. 
The $\tilde q_{i=1,..6}$, $\tilde\chi^{+}_{j=1,2}$ and $\tilde\chi^0_{l=1,..,4}
$ denote mass eigenstates of squarks, charginos and neutralinos, respectively. 
We calculate the Willson coefficients at the high energy ($m_W$) scale
for different contributions using the Feynman 
rules and notation presented in \cite{rosiek}
\begin{align}
\label{c7.mass}
(c_7)^{W^+}&=\frac{3}{2V_{cs}^*V_{us}}\sum_{q=d,s,b}V_{cq}^*V_{uq}\frac{m_q^2}{
m
_W^2}\bigl[e_d 
F_1\bigl(\tfrac{m_q^2}{m_W^2}\bigr)-F_2\bigl(\tfrac{m_q^2}{m_W^2}\bigr)\bigr]\\
(c_7)^{H^+}&=\frac{1}{2V_{cs}^*V_{us}}\!\!\sum_{q=d,s,b}\!\!V_{cq}^*V_{uq}\frac
{
m_q^2}{m_{H^+}^2}\bigl\{tg^2\beta\bigl[e_d 
F_1\bigl(\tfrac{m_q^2}{m_{H^+}^2}\bigr)-F_2\bigl(\tfrac{m_q^2}{m_{H^+}^2}\bigr)
\bigr]+\bigl[e_d 
F_3\bigl(\tfrac{m_q^2}{m_{H^+}^2}\bigr)-F_4\bigl(\tfrac{m_q^2}{m_{H^+}^2}\bigr)
\bigr]\bigr\}\nonumber\\
(c_7)^{\tilde 
g}&=-\frac{8~\alpha_s}{3~\alpha_WV_{cs}^*V_{us}}~e_u\sum_{i=1}^{6}\frac{m_W^2}{
m_{\tilde u_i}^2}\bigl[Z_U^{1i}Z_U^{2i*}F_2\bigl(\tfrac{m_{\tilde g}^2}{m_{\tilde u_i}^2}\bigr)-\frac{m_{\tilde g}}{m_c}Z_U^{1i}Z_U^{5i*}F_4\bigl(\tfrac{
m_{\tilde 
g}^2}{m_{\tilde u_i}^2}\bigr)\bigr]\nonumber\\
(c_7)^{\tilde 
\chi^+}&=-\frac{1}{V_{cs}^*V_{us}}\sum_{i=1}^{6}\sum_{j=1}^{2}\frac{m_W^2}{m_{
\tilde d_i}^2}\biggl\{a_{i1}^{j*}a_{i2}^j\biggl[- F_1\bigl(\frac{m_{\tilde 
\chi^{+}_j}^2}{m_{\tilde d_i}^2}\bigr)+e_dF_2\bigl(\frac{m_{\tilde 
\chi^{+}_j}^2}{m_{\tilde d_i}^2}\bigr)\biggr]\nonumber\\
&\qquad\qquad+\frac{m_{\tilde \chi^+_j}}{m_c}a_{i1}^{j*}b_{i2}^j\biggl[- 
F_3\bigl(\frac{m_{\tilde \chi^{+}_j}^2}{m_{\tilde 
d_i}^2}\bigr)+e_dF_4\bigl(\frac{m_{\tilde \chi^{+}_j}^2}{m_{\tilde 
d_i}^2}\bigr)\biggr]\biggr\}\nonumber\\
(c_7)^{\tilde 
\chi^0}&=-\frac{1}{V_{cs}^*V_{us}}~e_u\sum_{i=1}^{6}\sum_{l=1}^{4}\frac{m_W^2}{
m
_{\tilde u_i}^2}\biggl\{A_{i1}^{l*}A_{i2}^lF_2\bigl(\frac{m_{\tilde 
\chi^{0}_l}^2}{m_{\tilde u_i}^2}\bigr)+\frac{m_{\tilde 
\chi^0_l}}{m_c}A_{i1}^{l*}B_{i2}^lF_4\bigl(\frac{m_{\tilde 
\chi^{0}_l}^2}{m_{\tilde u_i}^2}\bigr)\biggr\}\nonumber
\end{align}
with  $e_d=-1/3$, $e_u=2/3$ and 
\begin{align}
\label{ab}
a^j_{iI}&=\frac{1}{g_2}\biggl[-\frac{e}{\sin\theta_W}Z_D^{Ii}Z_{1j}^-+\frac{\sqrt{2}m_d^I}{v_1}Z_D^{(I+3)i}Z_{2j}^-\biggr]V^{IJ*}\\
b^j_{iI}&=\frac{1}{g_2}\biggl[\frac{\sqrt{2}m_u^J}{v_2}Z_D^{Ii}Z_{2j}^{+*}\biggr
]V^{IJ*}\nonumber\\
A^l_{iI}&=\frac{1}{g_2}\biggl[-\frac{e}{\sqrt{2}\sin\theta_W\cos\theta_W}Z_U^{I
i
*}\bigl(\frac{1}{3}Z_N^{1l}\sin\theta_W+Z_N^{2l}\cos\theta_W\bigr)-\frac{\sqrt{
2
}m_u^I}{v_2}Z_U^{(I+3)i*}Z_N^{4l}\biggr]\nonumber\\
B^l_{iI}&=\frac{1}{g_2}\biggl[\frac{2\sqrt{2}e}{3\cos\theta_W}Z_U^{(I+3)i*}Z_N^
{
1l*}-\frac{\sqrt{2}m_u^I}{v_2}Z_U^{Ii*}Z_N^{4l*}\biggr]~.\nonumber
\end{align}
 The expressions for $c_7^\prime$ are analogous, but they are multiplied by an 
overall factor $m_u/m_c$ and $m_c$ is replaced by $m_u$ in (\ref{c7.mass}).  The matrices $Z^\pm$ are the mixing matrices of chargino mass eigenstates 
$\chi^+_{j=1,2}$ with weak eigenstates $(\tilde W^+,\tilde H^+)$ and the 
matrix $Z_N$ describes the mixing of the neutralino mass eigenstates 
$\chi^0_{l=1,..,4}$ with weak eigenstates $(\tilde B,\tilde A^3,\tilde 
H^0_1,\tilde H^0_2)$ \cite{rosiek}. The functions $F_{1,..,4}$ are given in 
Appendix B of \cite{bertolini}. The matrices $Z_{U,D}$ diagonalize the 
squark mass 
matrices and depend on a large number of unknown soft breaking parameters 
in the 
general MSSM. 

The number of these parameters entering the $c\to u\gamma$ amplitude is 
significantly reduced in the mass insertion approximation, where the 
expressions 
(\ref{c7.mass}) are expanded to the first order in the physical squark mass 
splitting  around the common average squark mass $m_{\tilde q}$ \cite{HF}, i.e.
\begin{equation}
f\biggl(\frac{m_{\tilde d_i}^2}{m_{\tilde \chi^+_j}^2}\biggr)\simeq 
f\biggl(\frac{m_{\tilde q}^2}{m_{\tilde \chi^+_j}^2}\biggr)+\frac{m_{\tilde 
d_i}^2-m_{\tilde q}^2}{m_{\tilde \chi^+_j}^2}f^\prime\biggl(\frac{m_{\tilde 
q}^2}{m_{\tilde \chi^+_j}^2}\biggr)~.
\end{equation}
This approximation works well \cite{bghw} unless there are strong 
cancellations 
or large flavour violations (for instance a light sbottom 
with a mass around $4$ GeV \cite{sbottom} could induce large flavor-violating
effects. However, its contribution to  $c\to u\gamma$ 
is negligible \footnote{The sbottom-chargino exchange gives 
$(c_7)^{\chi^+}\propto V_{cb}^*V_{ub}Z_{D}^{1i*}Z_{D}^{2i}~F(m_{\tilde 
\chi^+}^2/m_{\tilde b}^2)~m_W^2/m_{\tilde b}^2\propto 
V_{cb}^*V_{ub}Z_{D}^{1i*}Z_{D}^{2i}$ $m_{\tilde b}^2m_W^2/m_{\tilde \chi^+}^4$ 
(\ref{c7.mass}) in the limit $m_{\tilde b}^2/m_{\tilde \chi^+}^2\ll 1$, where the  unitarity of $Z_D$ has been applied.}). 

The summation  over the intermediate squarks can be expressed in terms of the 
squark mass matrices ${\cal M}_{\tilde U}^2$ and ${\cal M}_{\tilde D}^2$ 
in the 
super-CKM basis \cite{HF}. In this basis for the squarks,
the flavour structure of quark-squark-gaugino vertex is the same as the 
quark-quark-gauge boson vertex. We have
\begin{equation}
\label{sums1}
\sum_{i=1}^6Z_U^{Ii}Z_U^{Ji*}= \sum_{i=1}^6Z_D^{Ii*}Z_D^{Ji}=\delta^{IJ},\quad 
\sum_{i=1}^6Z_U^{Ii}Z_U^{Ji*}m_{\tilde u_i}^2=({\cal M}^2_{\tilde U})^{IJ},\quad 
 \sum_{i=1}^6Z_D^{Ii*}Z_D^{Ji}m_{\tilde d_i}^2=({\cal M}^2_{\tilde D})^{IJ}~.
\end{equation}
In evaluating (\ref{c7.mass}), we neglect the masses of the quarks
except $m_c$ because the magnetic transitions  ${\cal 
O}_7$ and ${\cal O}_7^\prime$ (\ref{c7.sm}) are of this order\footnote{We 
neglect the terms proportional to $m_u/m_c$ in $c_7^\prime$.}. The top
quark mass $m_t$ does not enter the amplitude and  
$m_b$ comes only in combination with a small factor $V_{cb}^*V_{ub}$. This 
approximation is supported by the fact that $c_7^{W^+}\propto V_{cq}^*V_{uq} 
m_q^2$ (\ref{c7.mass}) is about four orders of magnitude smaller than 
$(c_{7})_{\alpha_s}^{SM}$ (\ref{c7.sm}). Using the relations (\ref{sums1}) and 
the explicit expressions for the squark mass 
matrices\footnote{A similar procedure 
is used in the Appendix of \cite{HF} for $b\to s\gamma$.} we 
get\footnote{We omitt the terms proportional to $(M_{\tilde U}^2)_{RR}^{12}$ in $c_7^\prime$, 
since the best enhancement in general MSSM is expected when the upper bound on 
$\Delta m_D$ is saturated only by diagrams with with 
$(M_{\tilde U}^2)_{LL}^{12}$ insertions (see Table 1), while 
$(M_{\tilde U}^2)_{RR}^{12}$ is set to zero.}
\begin{align}
\label{c7}
(c_7)^{W^+}&,(c_7)^{H^+},(c_7^{\prime})^{W^+},(c_7^{\prime})^{H^+},(c_7^\prime)
^{\tilde \chi^+}\simeq 0\\
(c_7)^{\tilde 
g}&=\frac{8~\alpha_s~e_u}{3~\alpha_WV_{cs}^*V_{us}}~\frac{m_W^2}{m_{\tilde 
g}^4}~\biggl\{-(M_{\tilde U}^2)_{LL}^{12} f_2\bigl(\frac{m_{\tilde
 q}^2}{m_{\tilde g}^2}\bigr)+(M_{\tilde 
U}^2)_{LR}^{12}\frac{m_{\tilde g}}{m_c} f_4\bigl(\frac{m_{\tilde 
q}^2}{m_{\tilde g}^2}\bigr)\biggr\}\nonumber\\
(c_7)^{\tilde \chi^+}&=-\frac{1}{V_{cs}^*V_{us}}\sum 
_{j=1}^2\overbrace{[V(M_{\tilde 
D}^2)_{LL}V^\dagger]^{12}}^{\displaystyle{(M_{\tilde 
U}^2)_{LL}^{12}}}\frac{m_W^2}{m_{\tilde \chi_j^+}^4}\biggl\{|Z_{1j}^-|^2
\biggl[-f_1\bigl(\frac{m_{\tilde q}^2}{m_{\tilde \chi^+_j}^2}\bigr)+
 e_d f_2\bigl(\frac{m_{\tilde q}^2}{m_{\tilde 
\chi^+_j}^2}\bigr)\biggr]\nonumber\\
&\qquad\qquad\qquad\qquad\qquad-\frac{m_{\tilde 
\chi_j^+}}{\sqrt{2}m_W\sin\beta}Z_{1j}^{-*}Z_{2j}^{+*}
\biggl[-f_3\bigl(\frac{m_{\tilde q}^2}{m_{\tilde \chi^+_j}^2}\bigr)+
e_d f_4\bigl(\frac{m_{\tilde q}^2}{m_{\tilde \chi^+_j}^2}\bigr)\biggr]
\biggr\}\nonumber\\
(c_7)^{\tilde \chi^0}&=-\frac{e_u}{V_{cs}^*V_{us}}\sum 
_{l=1}^4\frac{m_W^2}{m_{\tilde 
\chi_l^0}^4}\bigl[\tfrac{1}{3}tg\theta_WZ_N^{1l*}+Z_N^{2l*}\bigr]\biggl((M_{\tilde U}^2)_{LR}^{12}\frac{m_{\tilde 
\chi_l^0}}{m_c}\biggl\{-\tfrac{2}{3}tg\theta_WZ_N^{1j*}f_4\bigl(\frac{m_{\tilde
 q}^2}{m_{\tilde \chi^0_l}^2}\bigr)\biggr\}\nonumber\\
&\qquad\qquad\qquad+(M_{\tilde 
U}^2)_{LL}^{12}\biggl\{\tfrac{1}{2}\bigl[\tfrac{1}{3}tg\theta_WZ_N^{1l}+Z_N^{2
l}\bigr]
f_2\bigl(\frac{m_{\tilde q}^2}{m_{\tilde \chi^0_l}^2}\bigr)+\frac{m_{\tilde 
\chi^0_l}}{2m_W\sin\beta}Z_N^{4l*}f_4\bigl(\frac{m_{\tilde q}^2}{m_{\tilde 
\chi^0_l}^2}\bigr)\biggr\}\biggr)\nonumber\\
%\end{align}
%\begin{align}
(c_7^{\tilde 
g})^\prime&=\frac{8~\alpha_s~e_u}{3~\alpha_WV_{cs}^*V_{us}}~\frac{m_W^2}{m_{\tilde g}^4}~(M_{\tilde U}^2)_{RL}^{12}~ \frac{m_{\tilde g}}{m_c} 
f_4\bigl(\frac{m_{\tilde q}^2}{m_{\tilde g}^2}\bigr)\nonumber\\
(c_7^\prime)^{\tilde \chi^0}&=-\frac{e_u}{V_{cs}^*V_{us}}\sum 
_{l=1}^4\frac{m_W^2}{m_{\tilde 
\chi_l^0}^4}\bigl[\tfrac{1}{3}tg\theta_WZ_N^{1l}+Z_N^{2l}\bigr](M_{
\tilde U}^2)_{RL}^{12}\frac{m_{\tilde 
\chi_l^0}}{m_c}\biggl\{-\tfrac{2}{3}tg\theta_WZ_N^{1j}f_4\bigl(\frac{m_{\tilde 
q}^2}{m_{\tilde \chi^0_l}^2}\bigr)\biggr\}\nonumber\\
{\rm with}\quad &\nonumber\\
f_{1,..,4}(x)&=\tfrac{\partial}{\partial 
x}\bigl[\tfrac{1}{x}F_{1,..,4}(\tfrac{1}{x})\bigr]_{x=0}~~.\nonumber 
\end{align}
The relevant diagrams for the various contributions to $c_7$ (\ref{c7}) in 
the mass 
insertion approximation are shown in Fig. 3. The expression $V(M_{\tilde 
D}^2)_{LL}V^\dagger$ occurring in the chargino contributions (\ref{c7}) 
is equal 
to $(M_{\tilde U}^2)_{LL}$  at the unification scale \cite{HF}. At the weak 
scale, this relation is valid up to the corrections proportional to quark 
masses, which can be  neglected within our approximation. The values of 
$c_7^\prime$ and the left-right induced $c_7$ are equal provided $|(M_{\tilde 
U}^2)^{12}_{LR}|\simeq |(M_{\tilde U}^2)^{12}_{RL}|$, which is indeed the case 
for the upper bounds given below (\ref{CD}).

\input{fig/graf3.tex}

QCD corrections will not drastically change our main result. 
As we will see, the contributions from charginos, 
neutralinos and left-left induced gluinos are generally smaller than those
in the standard model. Thus the main effect found in ref. \cite{greub},
namely that the corrections arising from ${\cal O}_2$ are dominant, remains
true for them. We need to include the QCD corrections only for the 
left-right induced gluino contribution because it might give a larger branching
 ratio 
than the $\simeq 3\times 10^{-8}$ of the standard model. Using
Eqs. (37) and (45) of ref. \cite{bghw}, one finds that the anomalous
dimensions of the relevant operators induce a $10 \%$ decrease of the 
amplitude 
(\ref{c7}), which is evaluated using $\alpha_s\!=\!\alpha_s(m_W)\!=\! 0.12$. \\

{\bf 3. The constraints on the mass insertions}

\vspace{0.2cm}

The Wilson coefficients  (\ref{c7}) depend on the mass insertions $(M_{\tilde 
U}^2)^{12}_{LL}$, $(M_{\tilde U}^2)^{12}_{LR}$ and $(M_{\tilde U}^2)^{12}_{RL}$
 in the super-CKM basis which incorporate the model dependence 
arising from the mechanism of supersymmetry breaking.  In  MSSM with  
universal soft breaking terms (i.e. models with gauge mediation 
\cite{gauge.mediation}) the squark mass matrices are flavour diagonal at the 
unification scale. Flavour mixing is due to the renormalization group 
evolution  down to the weak scale, because the  Yukawa matrices for up and down
 squarks are not simultaneously  diagonalizable.  The off-diagonal elements in 
squark mass matrices at the weak scale are necessarily proportional to the 
product of Yukawa couplings,   $(M_{\tilde U}^2)_{LL}^{12},~ (M_{\tilde 
U}^2)_{LR}^{12}\propto \sum _{q=d,s,b}V_{cq}^*V_{uq}m_{q}^2$ 
\cite{duncan.wyler}, and  the effect of this constrained  MSSM scenario 
on $c\to 
u\gamma$ decay rate is negligible. 

\vspace{0.2cm}

Supergravity scenarios \cite{supergravities} generally lead to the 
non-universal soft-breaking terms  and FCNC can be used to probe their 
textures. The mass insertions are taken as free parameters in this case, with 
bounds coming from available experimental data and the consistency of 
the model.  The mass insertions $(M_{\tilde U}^2)^{12}$  can be constrained by 
saturating the upper experimental upper limit on $\Delta m_D$, as done in 
\cite{gabbiani}. Taking the recent CLEO  result $(\Delta m_D\cos\delta +\Delta 
\Gamma_D\sin\delta/2)/(2\Gamma_D^2)<0.04\%$ \cite{CLEO} and assuming the 
relative strong phase $\delta$ between $D^0\to K^+\pi^-$ and $D^0\to K^-\pi^+$ 
to be small, the upper bounds on mass insertions are given in Table 1.
\begin{table}[h]
\begin{center}
\begin{tabular}{|c|c|c|}
\hline
 $m_{\tilde g}^2/m_{\tilde q}^2$ &  $|(M_{\tilde U}^2)_{LL}^{12}|/m_{\tilde 
q}^2$& $|(M_{\tilde U}^2)_{LR}^{12}|/m_{\tilde q}^2$\\
&  & $|(M_{\tilde U}^2)_{RL}^{12}|/m_{\tilde q}^2$\\
\hline
$0.3$&$0.03~m_{\tilde q}/500$ GeV&$0.04~m_{\tilde q}/500$ GeV\\
$1.0$&$0.06~m_{\tilde q}/500$ GeV&$0.02~m_{\tilde q}/500$ GeV\\
$4.0$&$0.14~m_{\tilde q}/500$ GeV&$0.02~m_{\tilde q}/500$ GeV\\
\hline
\end{tabular}
\caption{Limits on  $(M_{\tilde U}^2)_{LL}^{12}$, $(M_{\tilde U}^2)_{LR}^{12}$ 
and $(M_{\tilde U}^2)_{RL}^{12}$, obtained by saturating the upper bound 
$\Delta m_D<4.5\times 10^{-14}$ GeV of  CLEO \cite{CLEO}  by  gluino exchange 
\cite{gabbiani}.   }
\label{tab1}
\end{center}
\end{table} 

Stricter upper bounds on $(M_{\tilde U}^2)_{LR}^{12}$ and $(M_{\tilde 
U}^2)_{RL}^{12}$ are obtained by requiring the minima of the scalar potential 
in the MSSM not to break electric charge or color and that they are 
bounded from below \cite{CD}. In particular
\begin{equation}
\label{CD}
\frac{|(M_{\tilde U}^2)_{LR}^{12}|}{m_{\tilde q}^2}~,~\frac{|(M_{\tilde 
U}^2)_{RL}^{12}|}{m_{\tilde q}^2}\leq \sqrt{3}~ \frac{m_c}{m_{\tilde q}}\simeq 
 0.0023~\frac {500~{\rm GeV}}{m_{\tilde q}}~.
\end{equation}
The factor $m_c$ in (\ref{CD}) cancels the factor  $m_c$ in $(M_{\tilde 
U}^2)_{LR}^{12}/m_c$  (\ref{c7}), which in turn comes about since the chirality
 flip  in the  squark state does not have to be accompanied by the chirality 
flip in the external $c$ quark state. 

Another upper bound \cite{HF}
\begin{equation}
\label{relation}
(M_{\tilde U}^2)_{LL}^{12}\leq {\rm max}\biggl[(M_{\tilde 
D}^2)_{LL}^{12}~,~\frac{V^{12}}{V^{13}}(M_{\tilde U,\tilde 
D}^2)_{LL}^{13}~,~\frac{V^{12}}{V^{32}}(M_{\tilde U,\tilde 
D}^2)_{LL}^{32}\biggr]
\end{equation}
is due to the relation $(M_{\tilde U}^2)_{LL}=V(M_{\tilde D}^2)_{LL}V^\dagger$.
 Although the constraint on $(M_{\tilde D}^2)_{LL}^{12}$ from $\Delta m_K$ 
\cite{gabbiani} is about $30\%$ stronger than the constraint on $(M_{\tilde 
D}^2)_{LL}^{12}$, given in Table 1, the upper bound (\ref{relation}) is 
ineffective due to the large ratios $V^{12}/V^{32}$ and $V^{12}/V^{13}$.

We have explicitly verified that the upper bounds  on 
$(M_{\tilde U}^2)_{LL}^{12}$ 
(Table 1) and $(M_{\tilde U}^2)_{LR}^{12}$  (\ref{CD}) do not saturate the 
experimental value $(\Delta m_K)^{exp}\simeq 3.5\times 10^{-15}$ GeV through 
chargino exchange\footnote{The upper bound $(M_{\tilde U}^2)_{LR}^{12}$ 
(\ref{CD}) gives a $\Delta m_K$ which is several orders of magnitudes below 
$(\Delta m_K)^{exp}$. The upper bounds $(M_{\tilde U}^2)_{LL}^{12}$ (Table 
1)  can render values close to, but not above $(\Delta m_K)^{exp}$, for low 
masses of the charginos. The expressions for the chargino contributions to 
$\Delta m$ are presented in \cite{vives}. }. 
\\

{\bf 4. The results }

\vspace{0.2cm}

We consider low-energy supersymmetry with masses of superpartners in 
the TeV range. We take $250$ GeV$\leq m_{\tilde q},m
_{\tilde g}\leq 1000$ GeV with lower bounds given by the direct searches for 
these states \cite{PDG}. The masses of 
charginos, neutralinos and the corresponding mixing matrices $Z^\pm$, $Z_N$ can
 be expressed in terms of $tg\beta$, $\mu$, $m_1$ and $m_2$ \cite{rosiek,GH}. We take $tg\beta$ and $\mu$ in the range $2.5\leq tg\beta\leq 30$ and $100~{\rm 
GeV}\leq |\mu|\leq 300$ GeV. We assume $m_1\simeq m_2$ and determine this 
parameter by setting the mass of lighter chargino equal to the 
experimental lower bound $m_{\chi_1^+}>90$ GeV \cite{lepsusy} (we vary this 
parameter in the range $90-200$ GeV). Points of parameter space, 
where the neutralino masses violate bounds from the direct experimental 
searches, are omitted. 
The technical problem of negative eigenvalues for chargino and neutralino mass 
matrices is solved following the discussion in Appendix A.3 of \cite{GH}. The 
value $\alpha_s\!=\! \alpha_s(m_W)\!=\! 0.12$ is used.  

The predicted $c\to u\gamma$ branching ratios are given in Fig. 4 in a form of 
a 
scatter plot, with input parameters varying in ranges discussed above. The mass
 insertions $(M_{\tilde U}^2)_{LL}^{12}$  and $(M_{\tilde U}^2)_{LR,RL}^{12}$ 
are 
taken at the maximal values of the allowed ranges, given in Table 1 and  Eq. 
(\ref{CD}), respectively. The QCD-corrected standard model prediction 
(\ref{c7.sm}, \ref{br.sm}) is represented by the  dot on the left. The remaining dots represent branching ratios arising solely from the genuine supersymmetric contributions; as discussed above, QCD corrections  hardly affect these 
(\ref{c7}) and are neglected. The rates that would result from the 
gluino, the chargino and the neutralino  contributions (\ref{c7}) are shown 
separately.  If $(M_{\tilde U}^2)_{LR}^{12}$ is close to the upper value given 
by the vacuum stability argument (\ref{CD}), the  gluino exchange is 
dominated by the 
left-right mass insertion and  can enhance the standard model 
rate (\ref{br.sm}) by two orders of magnitudes    \begin{equation}
Br(c\to u\gamma)_{LR}\simeq 6\times 10^{-6}
\end{equation}
for $m_{\tilde q}\sim m_{\tilde g}\sim 250$ GeV. The process $c\to u\gamma$ is 
therefore particularly suitable for probing the universality of the trilinear 
soft-breaking $A$ terms, that are responsible for the left-right mass 
insertions.  If the dominant supersymmetric contribution comes from the 
$(M_{\tilde U}^2)_{LL}^{12}$ insertion (when $(M_{\tilde U}^2)_{LR}^{12}$ is 
negligible), then the gluino and chargino contributions are comparable in size 
and in fact add up. In this case,  the highest rate of genuine SUSY contribution  is achieved at small $tg\beta$ and $\mu<0$
\begin{equation}
Br(c\to u\gamma)_{LL}\simeq 10^{-9}~
\end{equation}
and would be screened by the standard model contribution (\ref{br.sm}). \\

\begin{figure}[!htb]
\begin{center}
\includegraphics[scale=1]{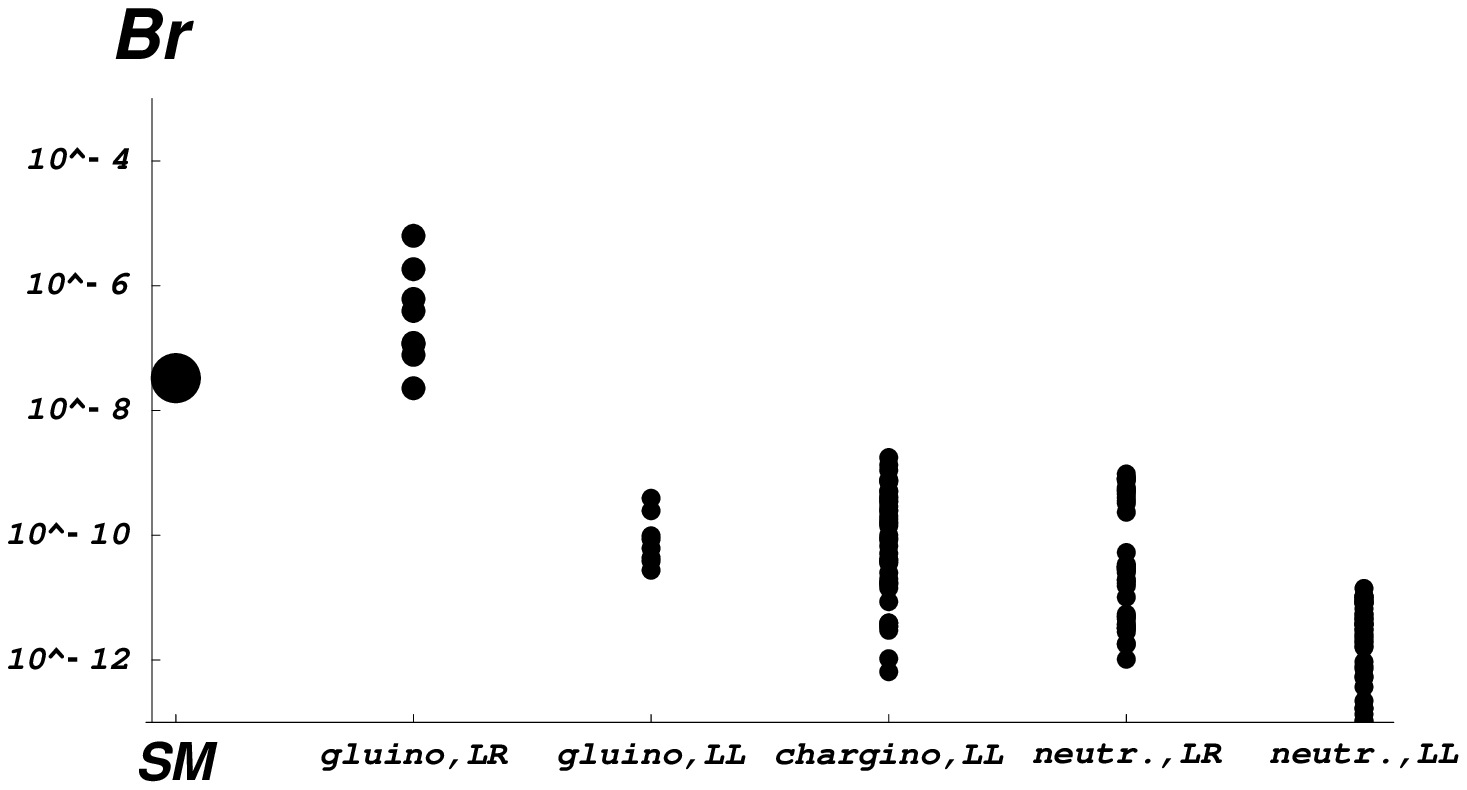} 
 \caption{The predicted $c\to u\gamma$ branching ratios for various contributions in MSSM.  
The QCD-corrected standard model rate (\ref{c7.sm}, \ref{br.sm}) \cite{greub} is represented by the  dot on the left. The remaining dots represent the rates arising solely from various supersymmetric contributions and do not include the QCD corrections (\ref{c7}): the gluino exchange (the contributions arising from $(M_{\tilde U}^2)_{LR,RL}^{12}$ and $(M_{\tilde U}^2)_{LL}^{12}$ are shown separately), the chargino exchange and the neutralino exchange (the contributions arising from $(M_{\tilde U}^2)_{LR,RL}^{12}$ and $(M_{\tilde U}^2)_{LL}^{12}$ are shown separately). The mass insertions $(M_{\tilde U}^2)_{LL}^{12}$  and $(M_{\tilde U}^2)_{LR,RL}^{12}$ are taken at the maximal values of the allowed ranges, given in Table 1 and  Eq. (\ref{CD}), respectively. Other parameters are varying in the ranges, discussed in the text. }  
\end{center}
\end{figure}

{\bf 5. Conclusions}

\vspace{0.2cm}

We have examined the sensitivity of the $c\to u\gamma$ decay rate in the
context of the minimal supersymmetric standard model. Not surprisingly,
schemes with universal soft breaking terms have a negligible effect on 
this process. If  the universality condition is relaxed, the chargino 
and neutralino contributions remain small and are 
at least one order of magnitude below the standard model rate.  However,
the gluino exchange can enhance the standard model rate 
of $Br(c\to u\gamma)\simeq 3\times 
10^{-8}$ by up to a factor of $200$, given the various constraints 
on the relevant left-right mass insertions. This is quite a surprising result
and makes this decay very attractive for flavor studies in supersymmetric
models. To understand this drastic effect, we recall that  the standard
model  branching ratio is essentially due to a two-loop diagrams in Fig. 1,
because the one loop-contribution is heavily GIM suppressed. As a consequence,
one may expect that the GIM non-suppressed two-loop 
amplitude is enhanced by a factor  
$\alpha_s(m_W)/(\alpha_WV_{cs}^*V_{us}\alpha_s(m_c))$ in a general MSSM, in 
qualitative 
agreement with our detailed numerical result. 

Although unwanted long distance effects strongly affect the hadronic rates, it 
is possible to find observables where they largely cancel. 
An important observable is $ R = [Br(D^0\to \omega\gamma)-Br(D^0\to 
\rho^0\gamma)]/Br(D^0\to \omega\gamma)$ which can go up to ${\cal O}(1)$ in
the general MSSM discussed here, as compared to a standard model prediction 
of $6\pm 15\%$ \cite{FPSW}. Although the long distance effects are
not so well under control \cite{pakvasa}, a value 
of $R$ above $20 \%$ would be a 
clear signal for new physics. Present experimental upper bounds on the rare 
decays $D^0\to 
\rho^0\gamma$ and $D^0\to 
\rho^0\omega$ are at the level 
of $10^{-4}$ \cite{CLEOrad,PDG}, while theoreticaly
predicted are at a level of $10^{-6}$ 
\cite{radiative}. It is expected that 
BaBar, Belle, BTeV  and possibly a tau-charm factory can
observe these interesting decays.

Another observable which is relatively free of long distance effects 
is $Br(B_c\to B_u^*\gamma)$. While  the standard model expectation is 
 $1\times 10^{-8}$ \cite{FPS}, it could reach $1\times 10^{-6}$ in
the MSSM. Such an  enhancement can be observed at LHC, which is expected to 
produce $2\times 10^{8}$ $B_c$ mesons with $p_T(B_c)>20$ GeV.\\

{\bf Acknowledgment}

\vspace{0.2cm}

We thank S. Pakvasa for a discussion on long distance effects. We would like to
 thank A. Masiero for many fruitful discussions and for reading the manuscript. We  thank D.A. Demir, O. Vives, E. Lunghi and S. Bertolini for useful 
discussions at the early stage of this work.

\end{document}

%% file: fig/graf1.tex
\begin{figure}[h]

\centering
\begin{fmffile}{zz1aaa}
  \fmfframe(3,3)(3,3){
  \begin{fmfgraph*}(30,30)
  \fmfpen{thin}
  \fmfleft{l3,l2,l1,l3}\fmfright{r4,r2,r1,r3}\fmftop{t1}\fmfbottom{b1}
  \fmfv{de.sh=square,de.filled=full,decor.size=4thick,label=${\cal O}_2^q=
   \bar u\gamma_{\mu}(1-\gamma_5)q~\bar q\gamma^{\mu}(1-\gamma_5)c$,la.d=4thick,la.a=90}{v}
  \fmf{plain,tension=1}{l1,al}\fmf{plain,tension=0.3}{al,v}
  \fmf{plain,tension=0.3}{v,ar}
  \fmf{plain,tension=1}{ar,r1}
  \fmffreeze
  \fmf{phantom,tension=2}{ml,l2}\fmf{phantom,tension=2}{mr,r2}
  \fmf{plain,left=0.4}{b,ml}
  \fmf{plain,right=0.4,label=$q=d,,s,,(b)$,la.si=right,la.d=1}{b,mr}
  \fmf{plain,left=0.5}{ml,v}\fmf{plain,right=0.5}{mr,v}
  \fmf{boson,tension=2}{b,b1}
  \fmffreeze
  \fmf{gluon,label=$g$,la.si=left}{ml,al}
  \fmflabel{$c$}{l1}\fmflabel{$u$}{r1}\fmflabel{$\gamma$}{b1}
  \end{fmfgraph*} }
\end{fmffile} 
\caption{The most important QCD corrections to $c\to u\gamma$  are induced by the four-quark operator ${\cal O}_2$ at the two loop level. Only one of the representative diagrams in the standard model is shown.}

\end{figure}

%% file: fig/graf2.tex
\begin{figure}[h]

\centering
\mbox{
\begin{fmffile}{zz2a}
  \fmfframe(1,1)(1,1){
  \begin{fmfgraph*}(30,25)
  \fmfpen{thin}
  \fmfleft{l1}\fmfright{r1}\fmftop{p1,t1,p3,p2,p4}
  \fmf{fermion,tension=1}{l1,v2}
  \fmf{fermion,tension=0.4,label=$d,,s,,b$,la.s=left}{v2,v3}
  \fmf{fermion}{v3,r1}
  \fmffreeze
  \fmf{zigzag,label=$W^+,,H^+$,tension=0.2,right=1}{v2,v3}
  \fmflabel{$c$}{l1}\fmflabel{$u$}{r1}
  \end{fmfgraph*} }
  \end{fmffile}
\quad
\begin{fmffile}{zz2b}
  \fmfframe(1,1)(1,1){
  \begin{fmfgraph*}(25,25)
  \fmfpen{thin}
  \fmfleft{l1}\fmfright{r1}\fmftop{p1,t1,p3,p2,p4}
  \fmf{fermion,tension=1}{l1,v2}
  \fmf{dashes,tension=0.4,label=$\tilde d_i$,la.s=left}{v2,v3}
  \fmf{fermion}{v3,r1}
  \fmffreeze
  \fmf{fermion,label=$\chi^+_j$,tension=0.2,right=1}{v2,v3}
  \fmflabel{$c$}{l1}\fmflabel{$u$}{r1}
  \end{fmfgraph*} }
  \end{fmffile}
\quad
\begin{fmffile}{zz2c}
  \fmfframe(1,1)(1,1){
  \begin{fmfgraph*}(30,25)
  \fmfpen{thin}
  \fmfleft{l1}\fmfright{r1}\fmftop{p1,t1,p3,p2,p4}
  \fmf{fermion,tension=1}{l1,v2}
  \fmf{dashes,tension=0.4,label=$\tilde u_i$,la.s=left}{v2,v3}
  \fmf{fermion}{v3,r1}
  \fmffreeze
  \fmf{fermion,label=$\tilde g,,\chi^0_l$,tension=0.2,right=1}{v2,v3}
  \fmflabel{$c$}{l1}\fmflabel{$u$}{r1}
  \end{fmfgraph*} }
  \end{fmffile}
       }
\caption{One-loop digrams for $c\to u\gamma$ decay in MSSM. Here $\tilde q_{i=1,..,6}$, $\chi^+_{j=1,2}$ and $\chi^0_{l=1,..,4}$ denote mass eigenstates for squarks, charginos and neutralinos. The photon is attached to any charged line.} 
\end{figure}

%% file: fig/graf3.tex
\begin{figure}[h]

\centering
\mbox{ 
\subfigure[Dominant gluino exchange diagrams. ]
{
\begin{fmffile}{q3a}
  \fmfframe(3,6)(3,6){
  \begin{fmfgraph*}(40,25)
  \fmfpen{thin}
  \fmfleft{l1}\fmfright{r1}\fmftop{t1,p1,p2,p3,p4}
  \fmf{plain,tension=1}{l1,v1}
  \fmf{plain,tension=0.5,label=$c_L$,la.s=right,la.d=20}{v1,v3}
  \fmf{dashes,tension=0.4,label=$\tilde c_L^\prime$,la.d=20,la.s=right}{v3,v4}
  \fmf{dashes,tension=0.4,label=$\tilde u_L^\prime$,la.d=20,la.s=right}{v4,v6}
  \fmf{plain,tension=1}{v6,r1}
  \fmffreeze
  \fmf{plain,label=$\tilde g$,la.s=right,right=1}{v3,v6}
  \fmflabel{$c_R$}{l1}\fmflabel{$u_L$}{r1}
  \fmfv{de.sh=pentagram,de.si=3thick,de.fill=full}{v1,v4}
  \fmfv{la.d=3thick,la.a=90,label=$m_c$}{v1}
  \fmfv{la.d=3thick,la.a=90,label=$(M_{\tilde U}^2)^{12}_{LL}$}{v4}
  \end{fmfgraph*} }
\end{fmffile}
\quad
\begin{fmffile}{qq3bb}
  \fmfframe(3,6)(3,6){
  \begin{fmfgraph*}(40,25)
  \fmfpen{thin}
  \fmfleft{l1}\fmfright{r1}\fmftop{t1,p1,p2,p3,p4}
  \fmf{plain,tension=1}{l1,v3}
  \fmf{dashes,tension=0.6,label=$\tilde c_R^\prime$,la.d=20,la.s=right}{v3,v5}
  \fmf{dashes,tension=0.6,label=$\tilde u_L^\prime$,la.d=20,la.s=right}{v5,v6}
  \fmf{plain,tension=1}{v6,r1}
  \fmffreeze
  %\fmf{boson,tension=2}{a,t1}
  \fmf{plain,label=$\tilde g$,la.s=right,right=1}{v3,v6}
  \fmflabel{$c_R$}{l1}\fmflabel{$u_L$}{r1}
  \fmfv{de.sh=pentagram,de.si=3thick,de.fill=full}{v5}
  \fmfv{la.d=3thick,la.a=90,label=$(M_{\tilde U}^2)^{12}_{LR}$}{v5}
  \end{fmfgraph*} }
\end{fmffile}
}
   }   
\mbox{ 
\subfigure[Dominant chargino exchange diagrams with flavour index $a,b=1,2,3$. ]
{
\begin{fmffile}{qqqq3c}
  \fmfframe(3,7)(3,7){
  \begin{fmfgraph*}(50,25)
  \fmfpen{thin}
  \fmfleft{l1}\fmfright{r1}\fmftop{t1,p1,p2,p3,p4}
  \fmf{plain,tension=1}{l1,v1}
  \fmf{plain,tension=0.5,label=$c_L$,la.s=right,la.d=20}{v1,v3}
  \fmf{dashes,tension=0.4,label=$\tilde d_{La}^\prime$,la.d=20,la.s=right}{v3,v4}
  \fmf{dashes,tension=0.4,label=$\tilde d_{Lb}^\prime$,la.d=20,la.s=right}{v4,v6}
  \fmf{plain,tension=1}{v6,r1}
  \fmffreeze
  \fmf{plain,label=$\tilde W^+$,la.s=right,right=1}{v3,v6}
  \fmflabel{$c_R$}{l1}\fmflabel{$u_L$}{r1}
  \fmfv{de.sh=pentagram,de.si=3thick,de.fill=full}{v1,v4}
  \fmfv{la.d=3thick,la.a=90,label=$m_c$}{v1}
  \fmfv{la.d=2thick,la.a=90,label=$V_{2a}^*$}{v3}
   \fmfv{la.d=2thick,la.a=90,label=$V_{1b}$}{v6}
  \fmfv{la.d=8thick,la.a=90,label=$(M_{\tilde D}^2)^{ba}_{LL}$}{v4}
  \end{fmfgraph*} }
\end{fmffile}
\quad
\begin{fmffile}{qqqqq3dd}
  \fmfframe(3,7)(3,7){
  \begin{fmfgraph*}(50,25)
  \fmfpen{thin}
  \fmfleft{l1}\fmfright{r1}\fmftop{t1,p1,p2,p3,p4}\fmfbottom{b1}
  \fmf{plain,tension=1}{l1,v3}
  \fmf{dashes,tension=0.8,label=$\tilde d_{La}^\prime$,la.d=20,la.s=right}{v3,v5}
  \fmf{dashes,tension=0.8,label=$\tilde d_{Lb}^\prime$,la.d=20,la.s=right}{v5,v6}
  \fmf{plain,tension=1}{v6,r1}
  \fmffreeze
  %\fmf{boson,tension=2}{a,t1}
  \fmf{plain,label=$\tilde H^+$,la.s=right,right=0.4,tension=0.7}{v3,vv}
  \fmf{plain,label=$\tilde W^+$,la.s=right,right=0.4,tension=0.7}{vv,v6}
  \fmf{phantom,tension=4}{vv,b1}
  \fmflabel{$c_R$}{l1}\fmflabel{$u_L$}{r1}
  \fmfv{de.sh=pentagram,de.si=3thick,de.fill=full}{v5}
  \fmfv{de.sh=circle,de.si=3thick,de.fill=full}{vv}
   \fmfv{la.d=2thick,la.a=90,label=$m_cV_{2a}^*$}{v3}
   \fmfv{la.d=2thick,la.a=90,label=$V_{1b}$}{v6}
  \fmfv{la.d=8thick,la.a=90,label=$(M_{\tilde D}^2)^{ba}_{LL}$}{v5}
  \end{fmfgraph*} }
\end{fmffile}
}
   }  
\mbox{ 
\subfigure[Dominant neutralino exchange diagrams. ]
{
\begin{fmffile}{q3e}
  \fmfframe(3,3)(3,3){
  \begin{fmfgraph*}(40,25)
  \fmfpen{thin}
  \fmfleft{l1}\fmfright{r1}\fmftop{t1,p1,p2,p3,p4}
  \fmf{plain,tension=1}{l1,v1}
  \fmf{plain,tension=0.5,label=$c_L$,la.s=right,la.d=20}{v1,v3}
  \fmf{dashes,tension=0.4,label=$\tilde c_L^\prime$,la.d=20,la.s=right}{v3,v4}
  \fmf{dashes,tension=0.4,label=$\tilde u_L^\prime$,la.d=20,la.s=right}{v4,v6}
  \fmf{plain,tension=1}{v6,r1}
  \fmffreeze
  \fmf{plain,label=$\tilde B,,\tilde A^3$,la.s=right,right=1}{v3,v6}
  \fmflabel{$c_R$}{l1}\fmflabel{$u_L$}{r1}
  \fmfv{de.sh=pentagram,de.si=3thick,de.fill=full}{v1,v4}
  \fmfv{la.d=3thick,la.a=90,label=$m_c$}{v1}
  \fmfv{la.d=3thick,la.a=90,label=$(M_{\tilde U}^2)^{12}_{LL}$}{v4}
  \end{fmfgraph*} }
\end{fmffile}
\quad
\begin{fmffile}{qq3ff}
  \fmfframe(3,3)(3,3){
  \begin{fmfgraph*}(40,25)
  \fmfpen{thin}
  \fmfleft{l1}\fmfright{r1}\fmftop{t1,p1,p2,p3,p4}\fmfbottom{b1}
  \fmf{plain,tension=1}{l1,v3}
  \fmf{dashes,tension=0.8,label=$\tilde c_L^\prime$,la.d=20,la.s=right}{v3,v5}
  \fmf{dashes,tension=0.8,label=$\tilde u_L^\prime$,la.d=20,la.s=right}{v5,v6}
  \fmf{plain,tension=1}{v6,r1}
  \fmffreeze
  %\fmf{boson,tension=2}{a,t1}
  \fmf{plain,label=$\tilde H_2^0$,la.s=right,right=0.4,tension=0.7}{v3,vv}
  \fmf{plain,label=$\tilde B,,\tilde A^3$,la.s=right,right=0.4,tension=0.7}{vv,v6}
  \fmf{phantom,tension=4}{vv,b1}
  \fmflabel{$c_R$}{l1}\fmflabel{$u_L$}{r1}
  \fmfv{de.sh=pentagram,de.si=3thick,de.fill=full}{v5}
  \fmfv{de.sh=circle,de.si=3thick,de.fill=full}{vv}
  \fmfv{la.d=2thick,la.a=90,label=$m_c$}{v3}
  \fmfv{la.d=3thick,la.a=90,label=$(M_{\tilde U}^2)^{12}_{LL}$}{v5}
  \end{fmfgraph*} }
\end{fmffile}
\quad
\begin{fmffile}{q3gg}
  \fmfframe(3,3)(3,3){
  \begin{fmfgraph*}(40,25)
  \fmfpen{thin}
  \fmfleft{l1}\fmfright{r1}\fmftop{t1,p1,p2,p3,p4}\fmfbottom{b1}
  \fmf{plain,tension=1}{l1,v3}
  \fmf{dashes,tension=0.8,label=$\tilde c_R^\prime$,la.d=20,la.s=right}{v3,v5}
  \fmf{dashes,tension=0.8,label=$\tilde u_L^\prime$,la.d=20,la.s=right}{v5,v6}
  \fmf{plain,tension=1}{v6,r1}
  \fmffreeze
  %\fmf{boson,tension=2}{a,t1}
  \fmf{plain,label=$\tilde B$,la.s=right,right=0.4,tension=0.7}{v3,vv}
  \fmf{plain,label=$\tilde B,,\tilde A^3$,la.s=right,right=0.4,tension=0.7}{vv,v6}
  \fmf{phantom,tension=4}{vv,b1}
  \fmflabel{$c_R$}{l1}\fmflabel{$u_L$}{r1}
  \fmfv{de.sh=pentagram,de.si=3thick,de.fill=full}{v5}
  \fmfv{de.sh=circle,de.si=3thick,de.fill=full}{vv}
  \fmfv{la.d=3thick,la.a=90,label=$(M_{\tilde U}^2)^{12}_{LR}$}{v5}
  \end{fmfgraph*} }
\end{fmffile}
}
   }  
\caption{The one-loop diagrams for $c\to u\gamma$ decay in MSSM within the mass insertion approximation. Only the diagrams, that give raise to the local operator ${\cal O}_7\propto \bar u_L\sigma^{\mu\nu} c_R$ (\ref{c7.sm}), are shown. Quark masses are neglected and only the terms of the order of ${\cal O}(m_c)$ are kept. The stars denote mass insertions for quarks and squarks. The squark mass insertions $ M_{\tilde U,\tilde D}^2$ and primed squark fields $\tilde q^\prime$ are represented in super-CKM basis. The dots denote mixing of different weak-isospin chargino or neutralino eigenstates. The photon is attached to any charged particle.}  
\end{figure}